\documentclass[a4paper,fleqn,usenatbib,useAMS]{mnras}

\usepackage{graphicx}    
\usepackage{amsmath}    
\usepackage{amssymb}    
\usepackage{multicol}        
\usepackage{bm}        
\usepackage{pdflscape}    
\usepackage{lineno}
\usepackage{appendix}

\usepackage[T1]{fontenc}
\usepackage{ae,aecompl}

\title[Automated detection of GRB]{A machine learning approach for GRB detection in \textit{AstroSat} CZTI data}

\author[S. Abraham et al]{Sheelu Abraham,$^{1,2}$\thanks{sheeluabraham@gmail.com}
Nikhil Mukund,$^{3,2}$\thanks{nikhil.mukund@aei.mpg.de}
Ajay Vibhute,$^{4,2}$
Vidushi Sharma,$^{2}$
\newauthor 
Shabnam Iyyani,$^{2}$
Dipankar Bhattacharya,$^{2}$
A. R. Rao,$^{5}$
Santosh Vadawale,$^{6}$
\newauthor 
Varun Bhalerao$^{7}$ \\
$^{1}$Marthoma College, Chungathara, Nilambur, Kerala. \\
$^{2}$Inter-University Center for Astronomy and Astrophysics, Post Bag 4, Ganeshkhind, Pune, India. \\
$^{3}$Max-Planck-Institut f{\"u}r Gravitationsphysik (Albert-Einstein-Institut) and Institut f{\"u}r Gravitationsphysik, \\ Leibniz Universit{\"a}t Hannover, Callinstra{\ss}e 38, 30167 Hannover, Germany  \\
$^{4}$Savitribhai Phule Pune University, Pune, Maharashtra, India. \\
$^{5}$Tata Institute of Fundamental Research, Mumbai, India. \\
$^{6}$Physical Research Laboratory, Ahmedabad, Gujarat, India. \\
$^{7}$Indian Institute of Technology, Bombay, India}

\begin{document}
\label{firstpage}
\maketitle

\begin{abstract}
We present a machine learning (ML) based method for automated detection of Gamma-Ray Burst (GRB) candidate events in the range 60 keV - 250 keV from the \textit{AstroSat} Cadmium Zinc Telluride Imager data. We use density-based spatial clustering to detect excess power and carry out an unsupervised hierarchical clustering across all such events to identify the different light curves present in the data. This representation helps understand the instrument's sensitivity to the various GRB populations and identify the major non-astrophysical noise artefacts present in the data. We use Dynamic Time Warping (DTW) to carry out template matching, which ensures the morphological similarity of the detected events with known typical GRB light curves. DTW alleviates the need for a dense template repository often required in matched filtering like searches. The use of a similarity metric facilitates outlier detection suitable for capturing previously unmodelled events. We briefly discuss the characteristics of 35 long GRB candidates detected using the pipeline and show that with minor modifications such as adaptive binning, the method is also sensitive to short GRB events. Augmenting the existing data analysis pipeline with such ML capabilities alleviates the need for extensive manual inspection, enabling quicker response to alerts received from other observatories such as the gravitational-wave detectors. 
\end{abstract}




\section{Introduction}

GRBs, the most energetic explosions known to occur, typically release $10^{46}$ - $10^{52}$ erg/s and last from a few milliseconds to a couple of minutes. Based on the duration over which $5\, \%$ to $95\, \%$ of the total burst fluence persists (T$\rm _{90}$), these events are often classified as short when T$\rm_{90}$ is less than $2$ s and long when it is otherwise \citep{1993ApJ_413L_101K,1995AAS_186_5301K}. The long GRB events are associated with the death of massive stars \citep{1993ApJ_405_273W,1998Natur_395_672I, macfadyen1999collapsars}, and this correlation has been confirmed with the coincident detection of a supernova 1c with the long GRB030329A \citep{2003ApJ_591L_17S}. The short GRBs are supposed to have a different progenitor and are likely to be produced due to the merger of compact objects like binary neutron stars or a neutron star and a black hole \citep{1989Natur_340_126E, 1992ApJ_395L_83N}. The recent discovery of gravitational waves from the binary neutron star merger GW170817 by the advanced \textit{LIGO} and advanced \textit{Virgo} observatories  \citep{2017PhRvL.119p1101A} together with the detection of a short GRB (GRB170817A) by various gamma-ray instruments such as {\it Fermi-GBM} and \textit{Integral} \citep{2017ApJ_848L_14G}, have confirmed this proposed mechanism, thus marking the beginning of the multi-messenger astronomy era. 

A GRB event can be divided into two main epochs: a prompt emission phase and a subsequent afterglow phase. The former occurs in gamma rays immediately after the initial burst trigger, while the latter is observed in multiple wavelengths from gamma rays to radio extending over a period lasting from days to months. Timely identification of prompt emission is necessary to carry out follow-up observation in multiple wavelengths by ground and space-based telescopes. This step can lead to the detection of afterglows, which is crucial in determining the GRB's redshift and various other properties. Simultaneous operation of multiple detectors capable of GRB detection would lead to improved sky coverage and constrain the event's time of occurrence to a higher degree of precision. Observing short-duration GRBs, in conjunction with a GW trigger, helps in understanding the kilonovae mechanisms.  Accurate time localization of these events can also constrain the differences in speed of light and gravity and thus scrutinize various theories of gravity \citep{ApJL_Abbott_2017}. 

The onboard alert systems of the Burst Alert Telescope (BAT; [50-150 keV]) on the \textit{Neil Gehrels Swift} Observatory \citep{2004ApJ...611.1005G} and of the Gamma-Ray Burst Monitor (GBM; [8 keV-40 MeV]) on the \textit{Fermi} satellite \citep{2009ApJ...702..791M}, have led to an increased number of detections along with more afterglow observations. Cadmium Zinc Telluride Imager (CZTI) onboard \textit{AstroSat}, is a wide field hard X-ray detector, and the increased transparency of collimators and surrounding supporting structures makes it sensitive to GRBs \citep{2016ApJ...833...86R}. In this paper, we overcome the absence of an onboard detector using an automated ML pipeline that enables low latency event detection in CZTI data. 

The paper is organized as follows: in Section~\ref{sec:data} the various pre-processing steps involved in generating light curves from CZTI data are presented. Section~\ref{sec:pipeline} briefly overviews the three machine learning algorithms used in this work and the proposed detection scheme. Section~\ref{sec:results} talks about the results from the blind search, while conclusions and prospects are presented in Section~\ref{sec:concl}.

\section{\textit{A\lowercase{stro}S\lowercase{at}} CZTI: Data and Preprocessing} \label{sec:data}
\textit{AstroSat} \citep{agrawal2006broad,2014SPIE.9144E..1SS} is India's first multi-wavelength space observatory capable of making observations in X-ray 
and UV bands. 
It carries the following five science instruments for simultaneous observations of the source of interest:  Ultra-Violet Imaging Telescope \citep[UVIT;][]{2017AJ....154..128T},  Large Area X-ray Proportional Counters \citep[LAXPC;][]{2016SPIE.9905E..1DY}, Soft X-ray Telescope \citep[SXT;][]{2017JApA...38...29S}, Cadmium Zinc Telluride Imager \citep[CZTI;][]{2017arXiv171010773R} and Scanning Sky Monitor \citep[SSM;][]{2017ExA....44...11R}. In particular, CZTI consists of an array of Cadmium Zinc Telluride (CZT) detectors, which are pixelated such that each pixel acts as an independent photon-counting detector. CZTI has a detector area of 976 cm$^2$ build using CZT modules and makes use of Coded Aperture Mask (CAM) for imaging \citep{bhalerao2017cadmium}. The total detection area is achieved by using 64 CZT modules of area 15.25 cm$^2$ each. These 64 modules are arranged in four identical and independent quadrants. The collimator walls separate these modules and collimators above each detector module, restrict the field of view to 4.6$^{\circ}$  x 4.6$^{\circ}$ (Full Width at Half Maximum) at photon energies below 100 keV. As the penetrating power of X-ray photons increases strongly with energy, the collimator slats, and the coded aperture mask transmits a significant fraction of photons above 100 keV, and the instrument behaves like an all-sky open detector enabling the detection of GRBs from any direction. 
It also carries a Caesium Iodide (Tl) based scintillator detector operating as anti-coincidence with the main CZT detector and is used as a veto detector. The coded aperture telescope is sensitive to hard X-ray polarization and was recently used to measure the polarized hard X-ray emission from Crab nebula \citep{2018NatAs...2...50V}. 

CZTI is configurable in 16 different modes. The default mode of operation is the event mode, denoted as Mode M0 (Normal Mode). CZTI also records accumulated spectral and housekeeping information once every 100 seconds and stores the recorded information when it is changed to Secondary Spectral Mode (Mode SS). Whenever the spacecraft passes through the South Atlantic Anomaly (SAA), High Voltage (HV) in the CZTI and Veto detectors are switched off, and the detector is in Mode M9 (SAA mode), during which only the housekeeping information is recorded once every second. During the normal mode, whenever a photon hits a detector, CZT records the photon's arrival time, its position on the detector plane, and the corresponding energy.  The time-tagged event list is stored in an event file. The events from four quadrants are stored as four different extensions of the event file. The recorded events also contain events generated due to the interaction of charged particles with the instrument or spacecraft body. The X-ray photons, consequently generated, can also deposit their energies in the CZTI detectors. Because of the pixelated nature of CZT, one charged particle can produce events in many pixels of CZT at the same time and are referred to as "bunches". During pre-processing, those bunches that do not belong to any astronomical source are mostly removed from the event file. Time intervals where data is not present due to SAA passage and data transmission errors are ignored, and a Good Time Interval (GTI) file is produced. The events belonging to GTIs are filtered and passed for further processing. During the data cleaning process, events from noisy or flickering pixels are also removed. The onboard calibration source, Am-241, emits X-rays of energy $60$ keV and an alpha particle simultaneously.  The alpha particle is absorbed in the CsI (TI) crystal, whereas the X-ray gets detected in the CZT pixel, and the alpha flag is set to 1. The events having the alpha flag equal to 1 are thereby removed from the event list. The cleaned event list so obtained is used as the input to the GRB detection algorithm. 

One event file from CZTI usually consists of multi-orbit data, which may span 6000 - 30,000 seconds. We have divided the data into small chunks of 500 seconds each to check for any trigger present. This step, however, limits the ability to identify a trigger that occurs between two such chunks of data. Each event file consists of data from all four quadrants and the four veto channels, and only those events with energy higher than 60 keV are considered for further analysis. We conduct the search in the count space, and the conversion from counts to flux depends on the effective area as a function of direction and energy. This conversion varies quite strongly for CZTI \citep{bhalerao2017cadmium} and requires prior knowledge of the transient location. For bright transients, there has been limited success in localising a burst \citep{Bhalerao_2017b}, and in such cases, one can attempt a joint solution for the source direction and spectrum. Consequently, it is possible that the event clustering is not related to the intrinsic characteristics of the GRB but only to the location in the relative detector coordinates. We also perform pre-processing to clean the time series before feeding it to the analysis pipeline. It is never perfect, and traces of SAA could still be visible in certain data segments. However, the pipeline is configured to take care of such noise sources, and in most cases, vetoes them successfully. 

\begin{figure} 
\begin{center}
\includegraphics[width=0.45\textwidth]{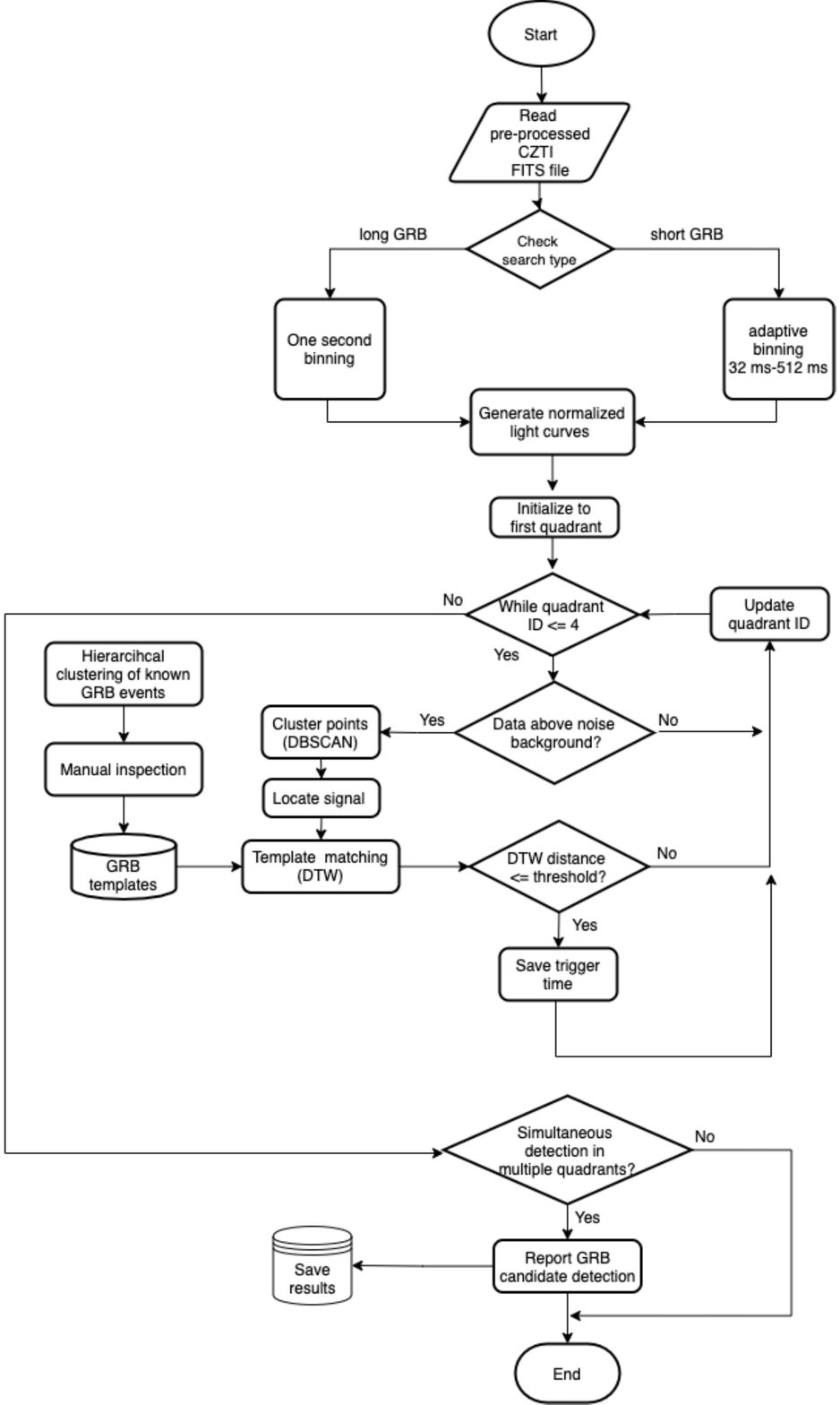}
\end{center}
\caption{Schematic depicting the steps involved in \textit{AstroSat} GRB candidate event detection pipeline.}
\label{fig:FC}
\end{figure}

\section{Candidate Event Selection} \label{sec:pipeline}
This section describes the framework (see Figure~\ref{fig:FC}) used for GRB candidate event detection and its resourcefulness in bringing down the time needed to issue Gamma-ray Coordinates Network (GCN) alerts to the broader astronomy community. We also discuss the different machine learning algorithms deployed to detect real GRBs candidates from false triggers that arise from artefacts such as instrumental noise, cosmic rays, or even random fluctuations. 

\subsection{Template Bank Generation}
\begin{figure*}
\begin{center}
\includegraphics[width=0.75\textwidth]{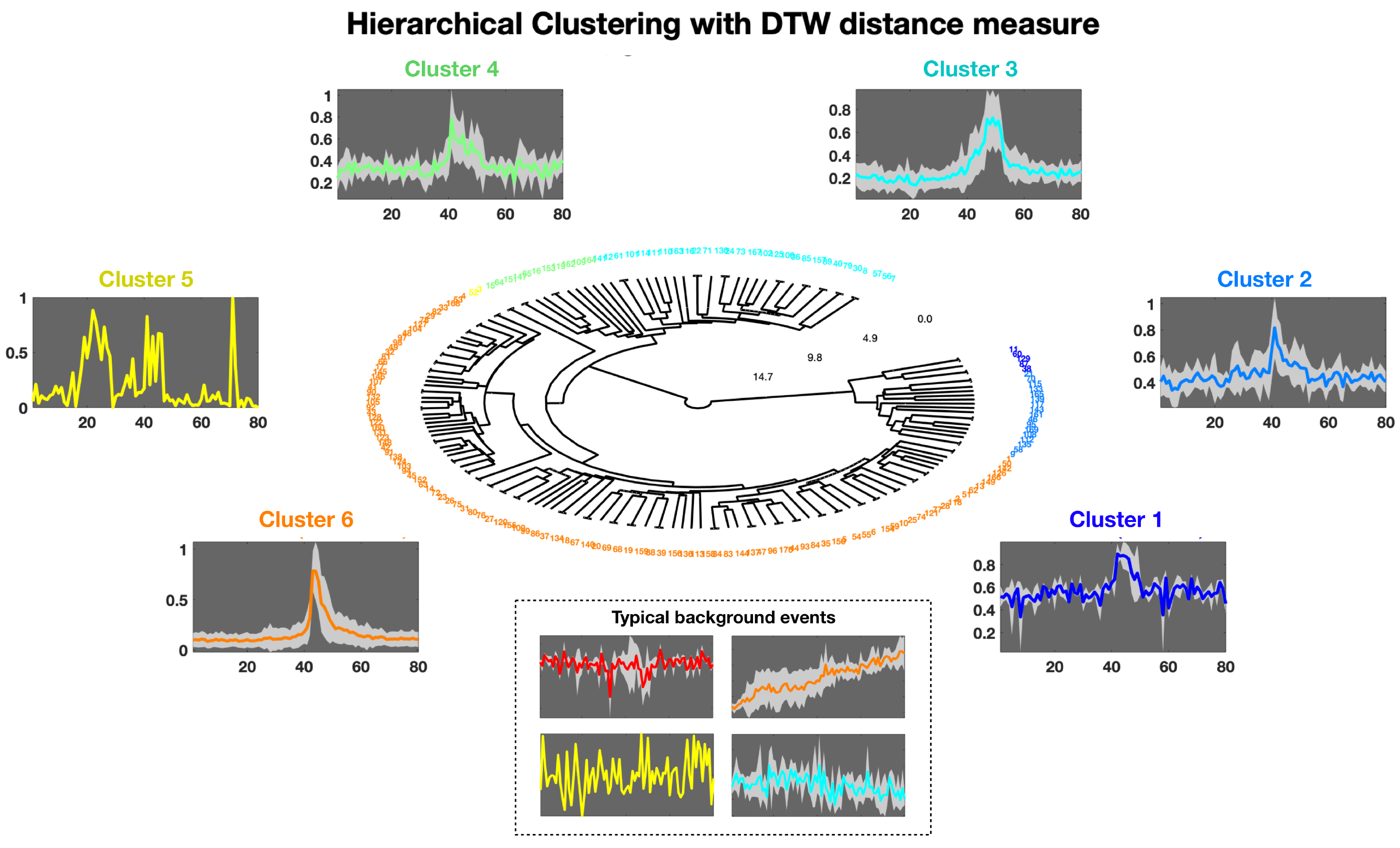}
\end{center}
\caption{Hierarchical clustering of known GRB light curves using DTW as the distance measure. Mean curve identified in each cluster is further utilized in the search for new events via DTW based template matching. Events within the box depict the typical background events seen in the CZTI data.}
\label{fig:HC}
\end{figure*}

We start with creating a template bank for long GRB light curves using 87 known GRBs\footnote{list of GRBs available as supplementary material}.
The key idea is to minimize the number of templates while still achieving maximal coverage of the light curves' morphologies. These templates are obtained from already identified CZTI data events using the GCN trigger information published by the currently operating space observatories. We carry out one second binning for each event and use the interval correlation optimized shifting (icoshift) technique \citep{savorani2010icoshift,tomasi2011icoshift} to correct for any delays among the light curves observed within the four quadrants. We then normalize every light curve individually by rescaling them to be within between 0 and 1. This process also brings down the mean DC background close to zero for all GRB-like events. These curves are further stacked together to get a temporal sequence with a higher signal-to-noise ratio (SNR) for each of the 87 known GRB events. We then carry out a hierarchical clustering using Dynamic Time Warping (DTW; See \ref{sssec:num1}) and identify the significant morphologies present within the data (see Figure~\ref{fig:HC}). The mean profile within each such cluster is then used to generate the GRB template bank. We use a bottom-up agglomerative clustering where the objects start as individual clusters, which are then hierarchically combined to form a dendrogram. The technique allows the user to choose any valid distance metric to compare the similarity between the objects. By maximizing the sum of similarities among the adjacent clusters, we can achieve optimal leaf ordering within the dendrogram \citep{doi:10.1093/bioinformatics/17.suppl_1.S22}. This ordering allows observing the progressively changing morphology within the given data samples. Hierarchical clustering has previously successfully identified the dominant groups among the short-duration transients, such as those observed in gravitational-wave observatories \citep{PhysRevD.95.104059}. 
We construct a template bank consisting of 52 GRB light curve templates based on the hierarchical clustering analysis results. We carefully choose these templates to guarantee adequate representation of all the probable morphologies of GRB events.

\subsection{Detection Scheme} \label{sssec:num1}

We perform one second binning for the 500s data chunks independently for each of the four quadrants.  We identify data significantly above the background noise by setting a threshold level three times the median absolute deviation above the median noise level. We then perform clustering using the DBSCAN \citep{ester1996density} algorithm on these samples and identify the significant temporal sequences that are later used in template matching analysis. DBSCAN, which stands for Density-based spatial clustering of applications with noise, groups together datasets with similar features and identifies outliers automatically. As compared to K-Means like clustering algorithms \citep{Hartigan1979}, there is no need to specify the number of clusters present in the data. The only required parameters for this algorithm are the minimum number of points in each cluster (\textbf{minPts}) and the maximum separation between samples (\textbf{eps}) for them to part of the same cluster. For this search looking for long GRB candidate events, we set to a value of five to both \textbf{minPts} and \textbf{eps}. We have used python implementation of DBSCAN from scikit-learn \citep{scikit-learn}.

\begin{figure*} 
\begin{center}
\includegraphics[width=0.8\textwidth]{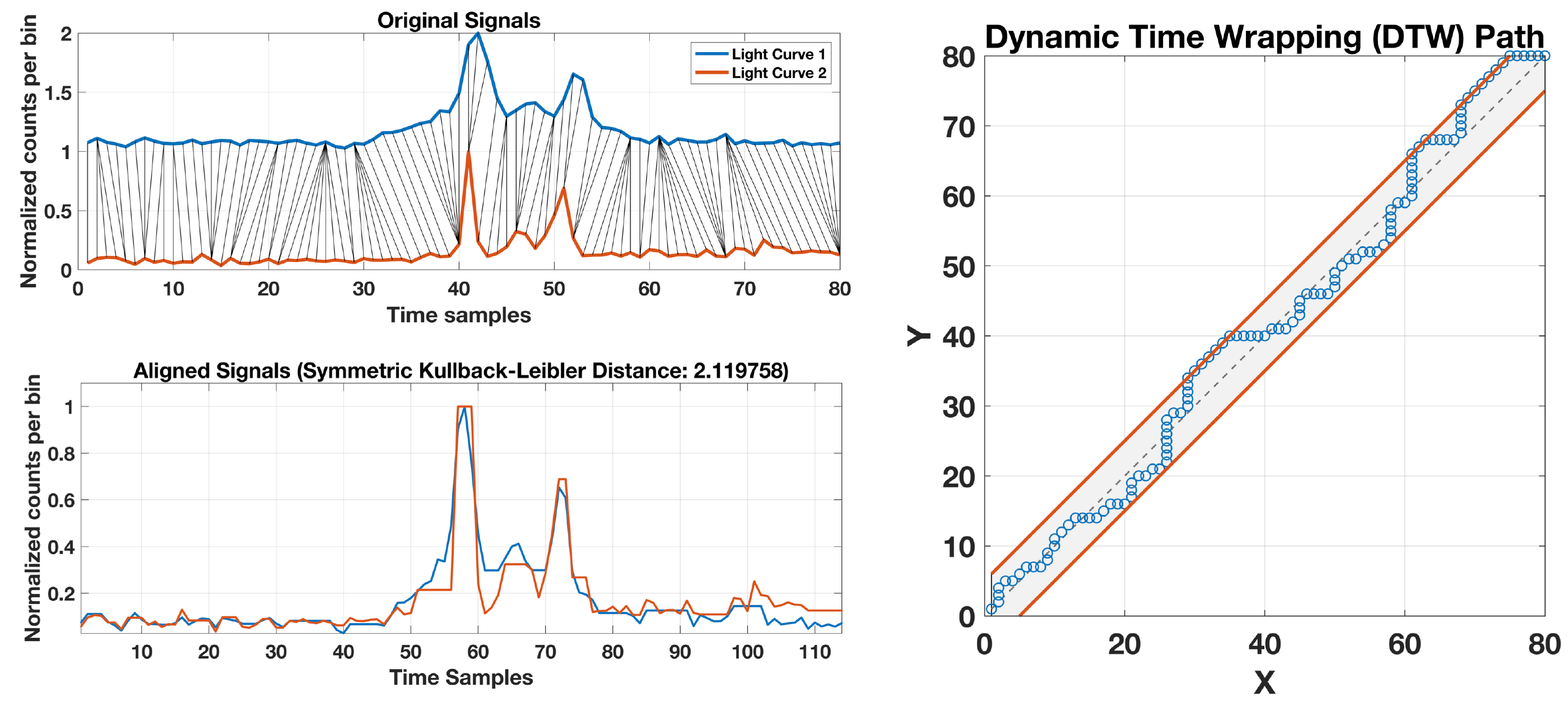}
\end{center}
\caption{Distance between two GRB light curves estimated using DTW with symmetric Kullback-Leibler metric and an adjustment window size of 5 samples. Each curve is individually normalized based on its peak height.}
\label{fig:DTW}
\end{figure*}

Once the clusters are identified, the corresponding temporal sequences are normalised and checked for similarity to known astrophysical signals using the above-mentioned DTW technique. It is a general method developed for time series alignment for speech and handwriting recognition. It can be applied to detect similar temporal sequences that are relatively stretched or squeezed with the template \citep{sakoe1978dynamic}. The method eliminates the need for feature extraction and can be easily extended to carry out the similarity search using a  template bank of known sequences. DTW finds the optimal alignment between time-series data, which allows a non-linear mapping of one signal to another, minimizing the distance between the two. To overcome the quadratic time and space complexity associated with the original DTW algorithm, we use FastDTW \citep{salvador2007toward} implementation, an approximation to DTW whose complexity is linear, thus speeding up the computation time. Pursuing alternative methods like cross-correlation or matched filtering in this scenario would require a dense template bank, making them computationally challenging for rapid detection. DTW has previously been demonstrated to be helpful in the similarity study between light curves from both GRBs and their X-ray flares \citep{Zhang_2016}. 

Let X and Y be two vectors of lengths M and N, respectively.  To create a mapping between the two vectors,  we need to define a path. The aim is to find the path of minimum distance. The optimal path starts from (0,0), ends at (M, N), and in between maps the vectors on to a common set of indices $i_x$ \& $i_y$ such that the total sum of distances, d
\begin{equation}
d = \sum_{\substack{m \;\epsilon\; i_x \\ n \;\epsilon\; i_y}} d_{m,n}(\textbf{X},\textbf{Y})
\end{equation}
is minimized where the distance $d_{m,n}$ is expressed in terms of symmetric Kullback-Leibler (KL) metric \citep{kullback1951},
 \begin{equation}
d_{m,n}(\textbf{X},\textbf{Y}) = (x_{m} - y_{n})(\log{x}_{m} - \log{y}_{n}) \; .
\end{equation}
The KL divergence is widely used in Bayesian inference and provides information about how well an approximate probability distribution represents the real underlying model.

The DTW path is constrained to move close to the diagonal by specifying a window around the main diagonal to minimize the effect of outliers. Additionally to ensure alignment of the complete signal and not just segments as well to prevent sample skipping, only the following transitions are permitted while the path proceeds from (0,0) to (M, N),
 \begin{eqnarray*}
(m,n) &\rightarrow& (m+1,n) \\
(m,n) &\rightarrow& (m,n+1) \\
(m,n) &\rightarrow& (m+1,n+1) \;.
\end{eqnarray*}

Figure~\ref{fig:DTW} shows one such instance of DTW-based alignment of two GRB light curves. 
We claim a detection if a trigger matches any of the template GRB models within a specified DTW distance and is coincidentally present in at least three quadrants channels.

\subsection{Performance Evaluation}
In general, we can assess the performance of the detection scheme described above from its receiver operator characteristic (ROC) curve, which compares the rate of detected actual events to the false triggers at varying detection threshold levels. Based on the available comparatively small data set, we carry out non-parametric modeling of both the GRB-like events and the expected noise sources and construct the ROC curves mentioned above. Combining GRNN with discrete wavelet transform has been previously shown to model temporal sequences effectively~\citep{WGRNN2011}. The usual discrete wavelet decomposition (DWT) splits the signal into approximation \& detail coefficients where the detail coefficients record the information lost between successive lower frequency approximations. However, wavelet packets decompose the details coefficients further into approximation and detail coefficients, thus improving the coarse resolution DWT and not being as computationally expensive as the continuous wavelet transform (CWT). We carry out such wavelet packet decomposition~\citep{Laine:1993:TCW:628307.628544,ta1994wavelet,walczak1996application} of the light curves using Daubechies wavelets and train a generalized regression neural network (GRNN) with these extracted features to generate the synthetic light curves. 

\begin{figure} 
\begin{center}
\includegraphics[width=1\linewidth]{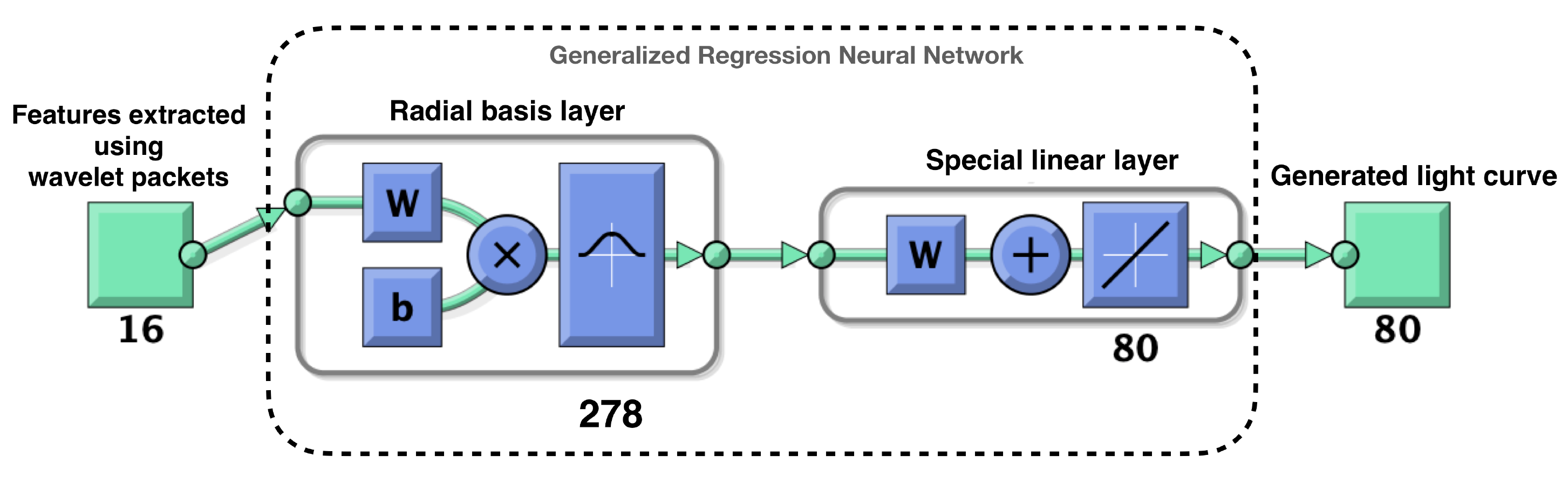}
\end{center}
\caption{Shallow two-layer generalized regression neural network with Gaussian kernels used to synthesize simulated light curves from a limited amount of training data. \textbf{W} represents the Weight, and \textbf{b} is the bias used in the respective neural network hidden layer. The number of radial basis layers (278) equals the number of training datasets. We use 80\% of light curves from 87 verified GRB events and use the data from all four quadrants.}
\label{fig:netGRNN}
\end{figure}

With their feed-forward shallow network architecture (see Figure~\ref{fig:netGRNN}), GRNNs avoid back-propagation and carry out a single pass learning with as many neurons as the number of datasets~\citep{Specht_GRNN_1991}. These networks use normalized radial basis function in their hidden layer, memorize all input-output sequences, and generalize them for newer inputs. These characteristics considerably decrease the overall training time and make them well suited for problems where training data availability is limited~\citep{959818}.  
We use the manually verified 87 GRB events and 36 non-astrophysical artefacts that include instances of SAA and cosmic rays to create a training dataset where 20\% is kept aside for validation. GRNN learning is, in general, sensitive to the variance of the involved radial basis function. We compare the network predictions against the validation data and optimize this parameter by minimizing the normalized mean squared error between the actual and predicted light curves. To generate synthetic events (1000 samples each for source and background events), we introduce random jitter at a few percent levels in the extracted wavelet parameters, draw samples from their distribution and feed them to the respective trained GRNNs. 

To access the relative improvement in performance, we compare the DTW classifier with a traditional peak finding algorithm on the synthetic dataset and depict the obtained ROC curves in Figure~\ref{fig:ROC}. These curves are constructed by varying the respective threshold parameter, DTW distance, and the peak prominence and calculating the true positive rate (TPR) and the false positive rate (FPR) at each of these points. As compared to the Euclidean metric, the DTW distance calculated using the symmetric KL metric provides better performance. We set the permissible FPR to 1\%  (TPR=98\%) and accordingly get an upper limit on the DTW threshold value to be 12 for the  KL metric-based distance estimation.
We carried out initial prototyping of the algorithms in MATLAB. The final detection pipeline is written in python for better integration with the rest of the satellite data analysis tools. The pipeline is currently configured to alert the CZTI \textit{AstroSat} support team about the most probable GRB candidates, who then makes the final decision on issuing GCN alerts. 

\section{Results} \label{sec:results}

\begin{figure} 
\begin{center}
\includegraphics[width=0.8\linewidth]{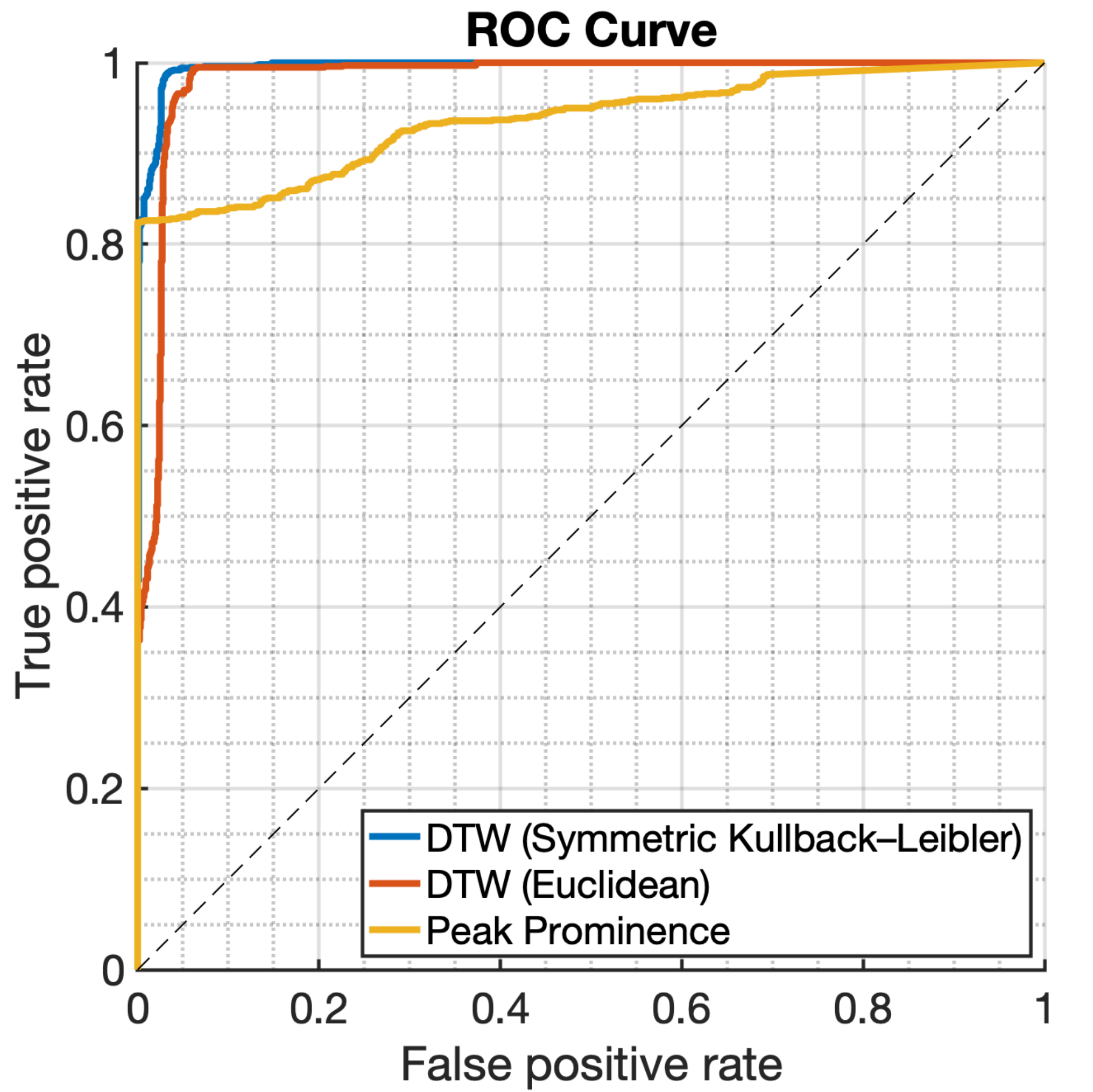}
\end{center}
\caption{Receiver Operator Characteristic (ROC) curve depicting the true positive rate vs. the false positive rate for both the DTW template matching scheme and a traditional peak detection algorithm. Curves are generated by varying the respective threshold parameter, DTW distance, and the peak prominence.}
\label{fig:ROC}
\end{figure}
%
%
%
\begin{figure*} 
\centering
\includegraphics[width=0.85\textwidth]{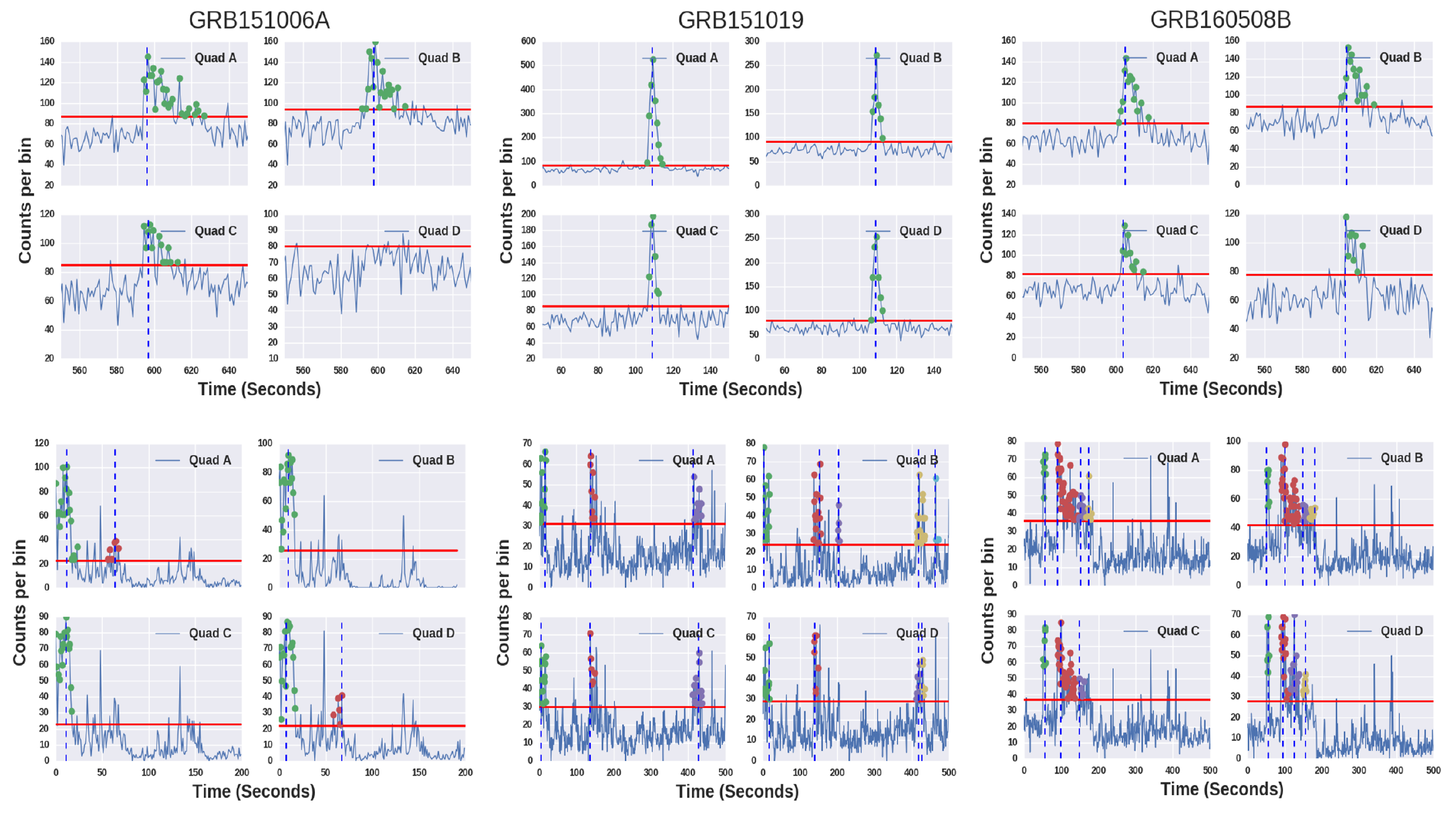}
\caption{Few detection scenarios encountered while searching for long GRBs with one-second binning. The upper panel shows the correctly identified events, while the lower panel highlights instances of contamination from non-GRB transients. The horizontal red line determines the threshold caused by the noise background level, and significant points above this are clustered using the DBSCAN algorithm. Each cluster is uniquely colored, with the vertical line providing the cluster center.}
\label{fig:fail}
\end{figure*}

Some of the detection scenarios involving true detection of long GRBs along with typical false-positive candidates are shown in Figure~\ref{fig:fail}. The upper panels show the correctly identified events, while the lower panels depicted certain instances when the events picked up were due to noise artefacts. After the testing and validation steps mentioned in the previous section, we carried out a blind search targeting long GRB events on the CZTI data collected from 8th October 2015 to 7th November 2018. The pipeline detected 223 probable candidates, out of which 170 were already known to be GRBs. Detailed analysis of the rest led to the discovery of 35 long GRBs candidate events along with 18 false positives. Table~\ref{tab:newgrb} lists these newly discovered events along with their trigger time in UTC, T$_{90}$\footnote{The calculation of T$_{90}$ and the other column values are described in Appendix A}, peak count rate, total count rate, mean background counts and detection significance with their uncertainties. Of these, GCN alerts were issued for two such events, namely
GRB180526A \citep{AstroSat_180526A} and GRB180603A \citep{AstroSat_180603A}. The observed number of false triggers are higher than the expected as shown in Figure~\ref{fig:ROC}. While the origin of many of these is not well understood, some of them seem to be related to the scenario shown in Figure~\ref{fig:fail} (lower right), where we see multiple short-duration transients all occurring within tens of seconds.  

\begin{figure*} 
\begin{center}
\includegraphics[width=\textwidth]{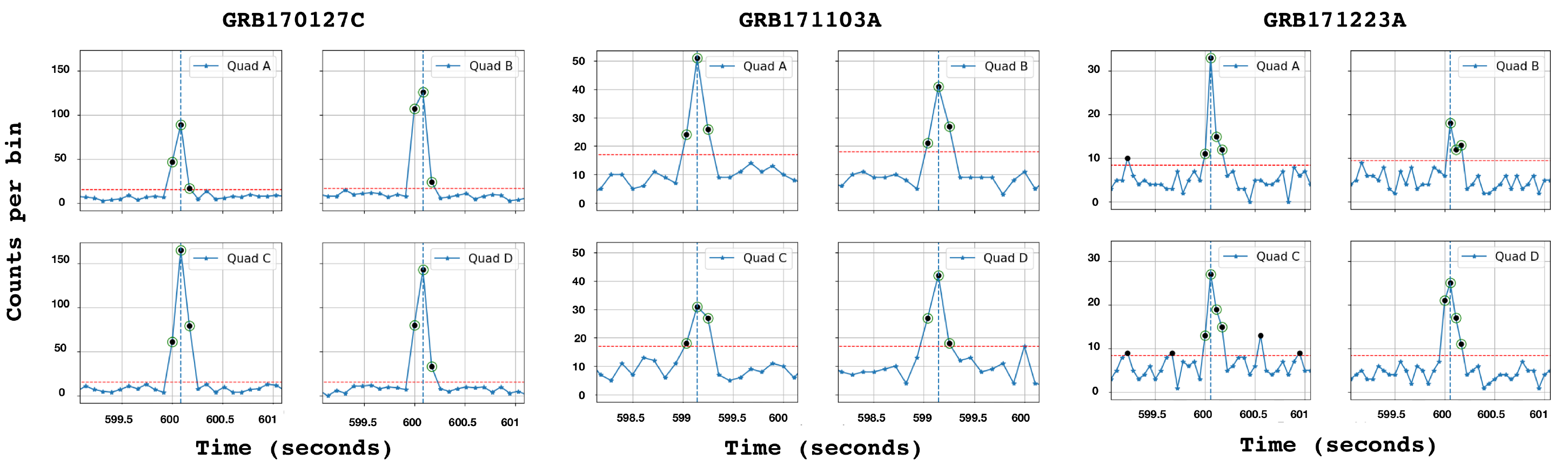}
\caption{Instances of short GRBs detected in CZTI data. In each case, we optimize the time binning to maximize the peak counts in multiple quadrants.  Time bins used for GRB170127C, GRB171103A, and GRB171223A are respectively 85ms, 106ms, and 55ms. The DBSCAN algorithm identifies significant clusters denoted by the green circles, followed by the DTW technique, which cross-matches them with the known GRB templates. }
\label{fig:shortGRB}
\end{center}
\end{figure*}

\vspace{0.1cm}
Certain minor modifications were necessary to make the pipeline sensitive to GRBs occurring at shorter timescales. As these events showed a higher sensitivity to the time bins' size, it was necessary to vary the bin size across timescales ranging from 32 ms to 512 ms for the entire data chunk and select the value that maximizes the peak count in at least three of the four quadrants. The time bin's optimal value was determined using the differential evolution \citep{storn1996usage} global optimizer, which performed a gradient-free direct search across this continuous parameter space. Besides, the $3\sigma$ detection threshold used for longer GRBs had to be increased to $4.5\sigma$ to minimize contamination from the background noise. 
 In Figure~\ref{fig:shortGRB} we report the triggers seen for GRB170127C \citep{Ajello_2019}, GRB171103A, and GRB171223A, which were respectively constructed using a bin size of 85ms, 106ms, and 55ms.

\begin{table*}
\begin{footnotesize}
\centering
\begin{tabular}{|c|c|c|c|c|c|c|c|}
 \hline & & & & &\\
  GRB ID& Trigger Time & T$_{90}$ & Peak   &  Total  & Mean  & Detection  \\
  & & & Count & Counts & Background & Significance  \\
  & (UTC) & (s) & Rate (s$^{-1}$) &  & Count Rate (s$^{-1}$) &   \\
 \hline & & &  & & &  \\
GRB151019A  & Oct 19 2015 8:05:25   & 6.1$\pm$0.16  & 1370$\pm$79.7  & 4877$\pm$112.4    & 453$\pm$25.2   & 64.4$\pm$4.2              \\
GRB151217A        & Dec 17 2015 4:58:21    & 29.6$\pm$0.07    & 108$\pm$30.0                 & 2030$\pm$137.1            & 503$\pm$22.6                      & 4.8$\pm$1.3               \\
GRB151219B         & Dec 19 2015 9:11:18   & 16.5$\pm$0.15    & 300$\pm$37.8                 & 1952$\pm$92.8             & 473$\pm$22.9                      & 13.8$\pm$1.8              \\
GRB151224A         & Dec 24 2015 2:26:35   & 7.1$\pm$0.19     & 117$\pm$30.3                 & 440$\pm$41.2              & 477$\pm$23.5                      & 5.4$\pm$1.4               \\
GRB160119C         & Jan 19 2016 3:08:26   & 31.2$\pm$0.12    & 233$\pm$31.8                 & 4819$\pm$147.2            & 423$\pm$21.5                      & 11.3$\pm$1.6              \\
GRB160128A         & Jan 28 2016 12:59:42  & 33.4$\pm$017     & 712$\pm$61.0                 & 9614$\pm$189.1            & 469$\pm$23.7                      & 32.9$\pm$2.9              \\
GRB160214A         & Feb 14 2016 9:17:26   & 26.6$\pm$0.1     & 526$\pm$55.7                 & 3840$\pm$139.1            & 463$\pm$30.8                      & 24.4$\pm$2.7              \\
GRB160221B         & Feb 21 2016 18:56:43  & 7.8$\pm$0.05     & 378$\pm$49.4                 & 1319$\pm$58.1             & 491$\pm$24.0                      & 17.1$\pm$2.3              \\
GRB160223B         & Feb 23 2016 9:59:03 & 13.5$\pm$0.15    & 239$\pm$38.5                 & 1985$\pm$85.3             & 480$\pm$28.8                      & 10.9$\pm$1.8              \\
GRB160310C        & Mar 10 2016 16:27:13  & 16.7$\pm$0.01    & 1481$\pm$82.9                & 13327$\pm$154.3           & 503$\pm$23.4                      & 66.0$\pm$4.0              \\
GRB160325B         & Mar 25 2016 6:59:23  & 43.6$\pm$0.02    & 1467$\pm$81.9                & 22263$\pm$279.8           & 474$\pm$22.8                      & 67.4$\pm$4.1              \\
GRB160418A         & Apr 18 2016 18:08:44  & 31.0$\pm$0.37    & 532$\pm$48.0                 & 7247$\pm$175.9            & 425$\pm$22.0                      & 25.8$\pm$2.4              \\
GRB160720A         & Jul 20 2016 18:25:23  & 14.5$\pm$0.06    & 430$\pm$41.8                 & 4244$\pm$99.9             & 459$\pm$23.7                      & 20.1$\pm$2.0              \\
GRB160805A         & Aug 05 2016 22:26:18  & 20.1$\pm$0.02    & 132$\pm$30.9                 & 218$\pm$60.3              & 428$\pm$22.3                      & 6.4$\pm$1.5               \\
GRB160824B        & Aug 24 2016 13:51:28  & 17.8$\pm$0.06    & 172$\pm$35.5                 & 2556$\pm$98.9             & 590$\pm$29.5                      & 7.1$\pm$1.5               \\
GRB160829B         & Aug 29 2016 14:18:47   & 18.0$\pm$0.05    & 1652$\pm$85.5                & 5438$\pm$131.5            & 504$\pm$25.0                      & 73.6$\pm$4.2              \\
GRB170210B         & Feb 10 2017 2:48:13   & 34.3$\pm$0.16    & 985$\pm$70.4                 & 13444$\pm$249.0           & 551$\pm$26.3                      & 42.0$\pm$3.2              \\
GRB170216A         & Feb 16 2017 16:39:33  & 14.8$\pm$0.19    & 301$\pm$41.0                 & 2634$\pm$91.0             & 508$\pm$24.0                      & 13.4$\pm$1.8              \\
GRB170228A         & Feb 28 2017 19:03:01  & 13.4$\pm$0.02    & 728$\pm$62.1                 & 4356$\pm$103.5            & 485$\pm$23.8                      & 33.1$\pm$2.9              \\
GRB170311C         & Mar 11 2017 13:45:10  & 7.4$\pm$0.05     & 486$\pm$53.9                 & 2152$\pm$59.1             & 517$\pm$23.8                      & 21.4$\pm$2.4              \\
GRB170316A         & Mar 16 2017 17:02:22  & 14.2$\pm$1.22    & 261$\pm$41.9                 & 1861$\pm$106.7            & 486$\pm$25.2                      & 11.8$\pm$1.9              \\
GRB170423B         & Apr 23 2017 20:55:22  & 12.9$\pm$0.05    & 441$\pm$46.6                 & 2413$\pm$77.1             & 474$\pm$23.6                      & 20.3$\pm$2.2              \\
GRB170614A         & Jun 14 2017 11:40:01  & 15.6$\pm$0.27    & 666$\pm$54.6                 & 6181$\pm$122.1            & 485$\pm$23.4                      & 30.2$\pm$2.6              \\
GRB170808B         & Aug 08 2017 22:27:47 & 12.1$\pm$0.05    & 1292$\pm$77.6                & 3679$\pm$79.3             & 460$\pm$22.4                      & 60.2$\pm$3.9              \\
GRB170825B         & Aug 25 2017 12:00:06  & 5.9$\pm$0.02     & 449$\pm$52.4                 & 1543$\pm$54.2             & 501$\pm$24.0                      & 20.1$\pm$2.4              \\
GRB170901B         & Sep 01 2017 11:59:57 & 11.4$\pm$0.04    & 310$\pm$43.3                 & 2138$\pm$66.7             & 495$\pm$23.3                      & 13.9$\pm$2.0              \\
GRB170915A         &  Sep 15 2017 3:51:28 & 10.2$\pm$0.07    & 246$\pm$38.7                 & 1708$\pm$72.9             & 543$\pm$25.4                      & 10.6$\pm$1.7              \\
GRB180401A         & Apr 01 2018 20:17:35   & 19.5$\pm$0.03    & 806$\pm$65.3                 & 6177$\pm$110.5            & 531$\pm$24.2                      & 35.0$\pm$2.9              \\
GRB180403A         & Apr 03 2018 13:32:52  & 7.0$\pm$0.06     & 186$\pm$38.0                 & 881$\pm$39.4              & 497$\pm$23.3                      & 8.3$\pm$1.7               \\
GRB180411C         & Apr 11 2018 12:28:32   & 75.0$\pm$0.02    & 360$\pm$47.9                 & 4603$\pm$222.6            & 487$\pm$24.1                      & 16.3$\pm$2.2              \\
GRB180416C         & Apr 16 2018 8:10:52 & 9.2$\pm$0.05     & 429$\pm$46.3                 & 2595$\pm$67.0             & 508$\pm$23.6                      & 19.0$\pm$2.1              \\
GRB180426A         & Apr 26 2018 13:10:59 & 12.4$\pm$0.03    & 434$\pm$49.7                 & 1944$\pm$61.6             & 498$\pm$23.3                      & 19.4$\pm$2.3              \\
GRB180504B         & May 04 2018 3:15:57  & 13.3$\pm$0.03    & 251$\pm$51.0                 & 1870$\pm$89.0             & 501$\pm$47.9                      & 11.2$\pm$2.3              \\
\textbf{GRB180526A}        & May 26 2018 11:04:18  & 57.8$\pm$0.15    & 541$\pm$55.6                 & 6504$\pm$213.9            & 497$\pm$24.6                      & 24.3$\pm$2.6              \\
\textbf{GRB180603A}        & Jun 03 2018 16:22:57  & 31.1$\pm$0.07    & 372$\pm$42.8                 & 6124$\pm$139.7            & 486$\pm$23.4                      & 16.9$\pm$2.0          \\

\hline 
\end{tabular}
\vspace*{0.2cm}
\caption{GRBs candidate events detected with the machine learning algorithm described in this paper. GCN circulars have been issued for the highlighted events.} \label{tab:newgrb}
\end{footnotesize}
\end{table*}

\section{Conclusions and outlook} \label{sec:concl}
We demonstrated various machine learning algorithms' resourcefulness for robust GRB detection using \textit{AstroSat} CZTI data. Automating such tasks can bring down the response time leading to efficient follow-up studies related to multi-messenger astronomy. Compared to conventional peak detection algorithms, incorporating morphology decreases the false detection rate from instrumental artefacts and non-GRB phenomena. The newly developed scheme has been tested on both short and long-duration GRB events and is now part of the \textit{AstroSat} CZTI data analysis pipeline. In the future, we would like to focus more on improving the detection efficiency for short GRBs through better time localization and a reduction in the number of false positives.  One natural way to achieve this would be to extend the template bank to include more models for real and spurious events. The techniques presented in this work are very well applicable to astronomical datasets such as stellar spectra or temporal sequences that are transient hence like the gravitational-wave transient signals. The feasibility of embedding such ML algorithms in FPGA-based hardware for low latency onboard trigger detection is also worth exploring in the context of next-generation detectors and would be part of future studies. 


\section{Acknowledgments}
The authors would like to thank the referees for their valuable comments, which helped improve the manuscript. This publication uses the data from the \textit{AstroSat} mission of the Indian Space Research Organisation (ISRO), archived at the Indian Space Science Data Centre (ISSDC). CZT-Imager is built by a consortium of institutes across India, including Tata Institute of Fundamental Research, Mumbai, Vikram Sarabhai Space Centre, Thiruvananthapuram, ISRO Satellite Centre, Bengaluru, Inter-University Centre for Astronomy and Astrophysics, Pune, Physical Research Laboratory, Ahmedabad, Space Application Centre, Ahmedabad: contributions from the vast technical team from all these institutes are gratefully acknowledged. The authors express thanks to the CZTI \textit{AstroSat} support cell at IUCAA for their help in data curation and pre-processing. NM acknowledges Council for Scientific and Industrial Research (CSIR), India, for providing financial support as Senior Research Fellow. The authors also express thanks to Ninan Sajeeth Philip for his valuable comments and suggestions. 

\section{Data Availability}
This publication uses the data from the \textit{AstroSat}  mission of the Indian Space Research Organisation (ISRO) and the data are  available in Indian Space Science Data Center (ISSDC) at \url{ https://astrobrowse.issdc.gov.in/astro_archive/archive/Home.jsp}.

\bibliographystyle{mnras}
\bibliography{Sheelu_GRB} 

\appendixtitleon
\appendixtitletocon
\begin{appendices}
\section{Light Curve Parameters}

The \textit{AstroSat} CZTI method of estimating the $T_{90}$ is based on the accumulation of counts. A similar process is used in BeppoSAX, HETE-2, CGRO/BATSE, and INTEGRAL observatories. At first, the light curve is generated with counts per time bin. We coarsely choose and store the information of the GRB that includes both pre and post background regions from the light curve. The count estimated in each time bin in the light curve has a Poisson error associated with it. We simulate 50,000 such light curves by randomly drawing each bin coutn from the corresponding Poissonian distribution.

For each simulated light curve, the parameters: $T_{90}$, peak count rate (PCR; which is the maximum count rate observed in the light curve of the GRB), accumulated total counts, and mean background rate are calculated as follows:

(a) The background is modeled by fitting the selected pre and post GRB background regions by a polynomial.  The local mean background rate (MBR) is obtained by averaging the count rates found in these regions, subsequently subtracted from the light curve. The peak count rate and the corresponding time found in the resultant light curve are noted. 

(b) Using the background-subtracted light curve, the cumulative counts per bin are plotted with time. The duration, $T_{90} $ is calculated as $T_{90} = T_{95} -  T_{5}$, where $T_{95}$ and $T_{5}$ are the times when $95\%$ and $5\%$  of the total GRB event counts are obtained respectively. The accumulated total counts in $T_{90}$ interval is also calculated.  We obtain the distribution for each parameter with the above steps, and its mean and standard deviation are used for reporting the parameter value and its uncertainty. In the case of count rate, we note that the standard deviation obtained from the distribution only reflects the variation arising from different simulations of the light curve, and therefore, the total error on the reported count rate value ($N/sec$) is the sum of the standard deviation of the distribution and the Poisson error ($\sqrt{N}$).  

The detection significance is calculated as $\rm PCR/\sqrt{MBR}$ and the error on it is obtained by the standard method of error propagation using the uncertainities reported for PCR and MBR. 
\end{appendices}

\end{document}